\documentclass[prl,a4paper,twocolumn,superscriptaddress,longbibliography]{revtex4-2}

\usepackage{amssymb}
\usepackage{amsmath}
\usepackage{amsfonts}
\usepackage{graphicx}
\usepackage{bm}
\usepackage[usenames,dvipsnames]{xcolor}
\usepackage{multirow}
\usepackage{comment}
\usepackage{pgfplots}
\usepackage{tikz}

\usepackage[colorlinks,citecolor=blue,linkcolor=red,urlcolor=blue]
{hyperref}

\usepackage[final]{pdfpages}
\usepackage{pgffor}

\DeclareMathOperator{\Tr}{Tr}

\DeclareMathOperator{\diag}{diag}

\newcommand{\sqHe}{{sqHe}}
\newcommand{\iqHe}{{iqHe}}
\newcommand{\NLSM}{{NL$\sigma$M}}

\newcommand{\TPT}{{TPT}}

\newcommand{\TD}{{2D}}

\makeatletter
\newcommand\footnoteref[1]{\protected@xdef\@thefnmark{\ref{#1}}\@footnotemark}
\makeatother

\makeatletter
\AtBeginDocument{\let\LS@rot\@undefined}
\makeatother

\begin{document}

\title{Bulk-edge correspondence at the spin-to-integer quantum Hall effect crossover in topological superconductors}

\author{M. V. Parfenov}

\affiliation{\mbox{L. D. Landau Institute for Theoretical Physics, Semenova 1-a, 142432, Chernogolovka, Russia}}

\affiliation{Laboratory for Condensed Matter Physics, HSE University, 101000 Moscow, Russia}

\author{I. S. Burmistrov}

\affiliation{\mbox{L. D. Landau Institute for Theoretical Physics, Semenova 1-a, 142432, Chernogolovka, Russia}}

\affiliation{Laboratory for Condensed Matter Physics, HSE University, 101000 Moscow, Russia}

\begin{abstract}
The spin and integer quantum Hall effects are two cousins 
of topological phase transitions in two-dimensional electronic systems. Their close relationship makes it possible to convert spin to integer quantum Hall effect by continuous increase in a symmetry breaking Zeeman magnetic field.  We study peculiarities of bulk-edge correspondence and a fate of massless edge and bulk topological (instantons) excitations at such the crossover in topological superconductors.
\end{abstract}

\date{version 5, \today }

\maketitle


Topological phase transitions ({\TPT}) are at a constant focus of physics research. Discovery of topological insulators and superconductors \cite{Schnyder2008,Schnyder2009,Kitaev2009,Chiu2016} gave a boost to research on {\TPT} in 
disordered electronic systems \cite{Altland2015,Morimoto2015,Meier2018,Karcher2019BDI,Antonenko2020,Haller2020,Bagrets2021,Shapiro2022,Kasturirangan2022,Matveeva2023,Antonenko2023,Kamenev2023,Altland2024,Zhao2024,Barkhofen2024}. Perhaps, the most famous example of the {\TPT} is the integer quantum Hall effect ({\iqHe}) in which different topological phases are labeled by $\mathbb{Z}$ (the set of the integer numbers). The {\iqHe} reflects an existence of the $\mathbb{Z}$-valued topological charge in two-dimensional ({\TD}) realization of class A in Altland-Zirnbauer symmetry classification of disordered Hamiltonians \cite{Wigner1951,Dyson1962a,Dyson1962b,Zirnbauer1996,Zirnbauer1997,Zirnbauer2005}. The {\iqHe}
has two close cousins in {\TD} topological superconductors which distinct 
topological phases  
are 
labeled by integers: the spin (class C) \cite{Volovik1997,Kagolovsky1999,Senthil1999} and thermal (class D) \cite{Senthil2000} quantum Hall effects.

The {\iqHe} has been investigated extensively in experiments \cite{Pruisken1988,Koch1991,Pruisken2000,Li2005,Pruisken2006,Li2009, Li2010,Shayegan2023,Kaur2024,Yeh2024} 
as opposed to
the spin quantum Hall effect ({\sqHe}). However, the latter has an advantage since its criticality is analytically tractable \cite{EversMirlin}. In particular, the position of the critical point \cite{Evers1997, Cardy2000}, the correlation length exponent, and the infinite subset of generalized multifractal exponents  are known exactly through the mapping to percolation \cite{GruzbergLudwig1999}. 
The class C can be thought as a parent class for the classes A and D  
due to  the following crossovers: C$\to$A with breaking of the SU(2) spin rotation symmetry down to U(1)~\footnote{Notably,  
there exists a reverse crossover from the class A to the class C controlled by the superconducting pairing \cite{Skvortsov2011}.} and C$\to$D which corresponds to the complete breaking of SU(2) symmetry while preserving superconductivity \cite{Kagolovsky1999,Senthil1999,GruzbergLudwig1999,Read2000,BhardwajKagalovsky2015,Kagalovsky2018}.

Although the crossover phenomena in the context of phase transitions are thoroughly studied \cite{Amit-book}, the crossovers between topologically non-trivial phases are much less investigated. An immediate difficulty can be readily appreciated from the observation that the topological phases of the {\sqHe} are enumerated by even integers  
 while the topological phases of the {\iqHe} are labeled by all integers. Thus the transformation $2\mathbb{Z}{\to}\mathbb{Z}$ should occur during the C$\to$A crossover. The understanding of crossovers between topological phases is complicated by the presence of topological excitations (instantons) in the bulk and massless edge excitations which are related by the bulk-boundary correspondence. 
From practical point of view, interest to the crossovers lies in their potential experimental applications. For instance, does the {\iqHe} realized in a topological superconductor due to the C$\to$A crossover differ from the ordinary {\iqHe} experimentally?

The goal of this Letter is to study the {\sqHe}-to-{\iqHe} crossover and answer the following 
{\it physical} questions:
(i) Is it possible to describe the crossover in terms of the edge theory only?
(ii) How do physical observables depend on a bare spin Hall conductance after the crossover? 
(iii) What is the structure of the emergent {\iqHe} staircase?

\noindent\textsf{\color{blue} Edge modes for {\sqHe}.} Both sqHe and iqHe possess non-dissipative gapless edge modes. We start from discussion of their transformation across the crossover. 
We begin with a reminder of the edge theory for the {\sqHe}~\cite{Senthil1999}. We consider chiral fermion quasiparticles at the edge of a ({\TD}) disordered $d_{x^2{-}y^2}{+}i d_{xy}$ superconductor. To be able to average over quenched disorder we will use the replica trick. The imaginary time replica action for the  {\sqHe} edge can be written in terms of $N_r$-copies of spin $1/2$ chiral fermions \cite{Senthil1999}:
\begin{equation}\label{eq: psiact1}
    S_{\text{e}} {=} \!\!\int\limits_{0}^{\beta}\!\! d\tau\!\!\! \int\!\! dy \Bigl [ \bar\psi(iv \partial_y {-}\partial_\tau{-}\eta_3)\psi {+} \eta_{-} \bar{\psi} \Sigma_{+} \bar{\psi}^{\rm T}{+}\eta_{+} \psi^{\rm T} \Sigma_{-}\psi\Bigr ] .
\end{equation}
Here $\bar{\psi}{=}\{\bar{\psi}_{\uparrow,1},{\dots},\bar{\psi}_{\downarrow,N_r}\}$ and ${\psi}{=}\{{\psi}_{\uparrow,1},{\dots},{\psi}_{\downarrow,N_r}\}^T$ are Grassmann variables corresponding to fermion creation and annihilation operators, $\Sigma_{\pm}{=}\sigma_{\pm}{\otimes} 1_r$ with $1_r$ being the identity matrix in the replica space and  
$\sigma_{\pm} {=} (\sigma_{1} {\pm} i \sigma_{2} )/2$ where $\sigma_{j}$ are standard Pauli matrices acting in the spin space. A quasiparticle edge velocity is denoted as $v$, and $\beta$ stands for the inverse temperature. The random Gaussian fields $\eta_{\pm}{=}\eta_1{\pm} i \eta_2$ and $\eta_3$ mimic fluctuations of a superconducting order parameter and scattering off impurities, respectively.  
They have the zero mean and  
are delta-correlated in space: $\langle \eta_{j}(y) \eta_{k}(y^{\prime}) \rangle {=} \varkappa\delta_{jk}  \delta(y{-}y^{\prime})$. 
The action~\eqref{eq: psiact1} does not conserve the number of $\psi$-fermions but has SU$(2)$ 
symmetry corresponding to spin conservation.

To elucidate symmetries of action \eqref{eq: psiact1} inherent in the class C, we introduce new fields: $\bar{\chi}_{\uparrow,\alpha}{=}\bar{\psi}_{\uparrow,\alpha}$, $\chi_{\uparrow,\alpha}{=}\psi_{\uparrow,\alpha}$, $\bar{\chi}_{\downarrow,\alpha}{=}\psi_{\downarrow,\alpha}$, and $\chi_{\downarrow,\alpha}{=}\bar{\psi}_{\downarrow,\alpha}$, where $\alpha{=}1,\dots,N_r$ \cite{Senthil1999}. In this representation the edge action \eqref{eq: psiact1} becomes
\begin{equation}
   S_{\text{e}} {=} \int_{0}^{\beta} d\tau\!\! \int dy \bar{\chi}({-}\partial_\tau {-} H{\otimes} 1_{r})\chi, \quad H {=} {-}i v \partial_y {+} \boldsymbol{\eta} \boldsymbol{\sigma} .
   \label{eq:H:chi}
\end{equation}
The above action conserves the number of $\chi$ fermions which coincides with the $z$-projection of the spin of $\psi$-fermions.
Thus, $\chi/\psi$-fermions are spin/charge carriers.   
Hamiltonian~\eqref{eq:H:chi} manifests 
anti-unitary Bogoliubov-de Gennes (BdG) symmetry, $H {=} {-}\sigma_2 H^T \sigma_2$, 
as expected for the class C.
The action~\eqref{eq:H:chi} describes two spin-degenerate hybridized electron-hole edge modes that propagate in  the 
same direction and transfer a quantum of the transverse spin conductivity each  \cite{Senthil1999}. Therefore, in the case of a clean system, (for which $\eta_j{\equiv}0$ and $N_r{=}1$), applying a generalized TKNN formula \cite{TKNNform},  we obtain that the spin Hall conductance is quantized in units of $2e^2/h$ \cite{Volovik1997,Senthil1999}
\begin{equation}
    g_{\rm H} {=} 2k (e^2/h) . 
    \label{eq:gH:sqHe}
\end{equation}
Here $k$ is the number of edge modes ($k{=}1$ for Eq. \eqref{eq:H:chi}).

\noindent\textsf{\color{blue} Edge theory for {\sqHe}.}\! As expected, the $2\mathbb{Z}$ quantization of $g_{\rm H}$, Eq.~\eqref{eq:gH:sqHe}, holds in the presence of the disorder. Averaging action \eqref{eq: psiact1} over disorder and employing the non-Abelian bosonisation \cite{witten1984non,Nersesyan1994,NERSESYAN1995561,ALTLAND2002283,Altlandgraph2006,KonigTop2013,KonigTop2014}, we derive the nonlinear sigma model ({\NLSM}) action for the soft diffusive edge modes (see Supplemental Materials \cite{SM}):
\begin{equation}\label{eq:Qedgeact}
    \mathcal{S}_{\rm e} {=} \frac{k}{2}\Tr \Lambda T\partial_y T^{-1} {+} \pi k \nu_{\text{e}} \Tr\hat{\epsilon} Q
        {-}\frac{k v^2}{16\varkappa} \Tr\left(\partial_y Q \right)^2  .
\end{equation}
Here $\nu_{\text{e}} {=} 1/(2 \pi v)$ is the density of edge states. $Q{=} T^{{-}1}\Lambda T$ is Hermitian traceless matrix acting in $N_r {\times} N_r$ replica, $2N_m {\times} 2N_m$ Matsubara frequencies, and  $2{\times}2$ Nambu spaces. The matrix $Q$  satisfies the following relations 
\begin{equation}\label{eq:Qprop}
    Q^2 = 1, \quad Q = Q^\dag = -L_0 \textsf{s}_2 Q^{T} \textsf{s}_2 L_0 .
\end{equation}
Here and below $\textsf{s}_{0,1,2,3}$ stand for the standard Pauli matrices in the Nambu space, $(L_0)_{nm}^{\alpha_1\alpha_2}{=} \delta_{\varepsilon_n,-\varepsilon_m} \delta^{\alpha_1 \alpha_2}\textsf{s}_0$, $\Lambda_{nm}^{\alpha_1 \alpha_2} {=} \text{sgn}(\varepsilon_n) \delta_{nm} \delta^{\alpha_1 \alpha_2} \textsf{s}_0$, $\hat\epsilon_{nm}^{\alpha_1 \alpha_2} {=} \varepsilon_n \delta_{nm} \delta^{\alpha_1 \alpha_2} \textsf{s}_0$, where $\varepsilon_n{=}\pi(2n{+}1)/\beta$ denotes the fermionic Matsubara frequency. Symbol `$\Tr$' includes spatial integration as well as the trace over replica, Matsubara, and Nambu spaces. 
As the consequence of SU(2) symmetry, the {\it spin space} 
is {\it not present} in action \eqref{eq:Qedgeact}. The relations \eqref{eq:Qprop} determine the {\NLSM} target manifold of class C, $Q{\in}{\rm Sp}(2N)/{\rm U}(N)$, where $N{=}2N_r N_m$, whereas $T{\in}{\rm Sp}(2N)$.  

The information about the quantization of $g_{\rm H}$ is encoded in the first term of the {\NLSM} \eqref{eq:Qedgeact}, which is nothing but the  
edge form of the 
Pruisken's $\theta$-term \cite{pruisken1984localization}. The factor $k/2$ is responsible for 
the exactly same result for $g_{\rm H}$ as in the clean case, Eq. \eqref{eq:gH:sqHe}. It is expected since  the gauge transformation 
$\Tilde{\chi}(y) {=} \mathcal{T}_y \exp [i \int^y dy^\prime \boldsymbol{\eta}(y^\prime) \boldsymbol{\sigma}/v_{\rm e}]\chi(y)$ ($\mathcal{T}_y$ is spatial ordering) \cite{Senthil1999} excludes disorder from  
Eq.~\eqref{eq:H:chi}.

\begin{figure}
    \centering
    \includegraphics[width=0.8\columnwidth]{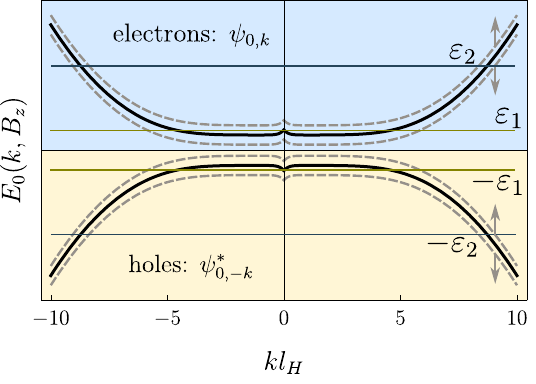}
    \caption{The quasiparticle spectrum in the toy-model: 2D fermions on a stripe in the presence of a perpendicular magnetic field, constant superconducting pairing amplitude, and Zeeman splitting (see \cite{SM} and Ref. \cite{antonenko2024,Kita2002,Zhao2025} for details).}
    \label{pic:spec1}
\end{figure}

\noindent\textsf{\color{blue} The {\sqHe}-to-{\iqHe} crossover at the edge.} In order to remove the spin degeneracy of the chiral edge states, 
we introduce the Zeeman magnetic field $B_z$ by adding the term  $\mu_{\rm B} B_z \int d\tau dy \bar{\psi}(\sigma_3{\otimes}1_r)\psi$ to the action \eqref{eq: psiact1}. Here $\mu_{\rm B}$ is the Bohr magneton. In Eq.~\eqref{eq:H:chi}, it works as 
the shift $H{\to}H{+}\mu_{\rm B} B_z\sigma_0$, that explicitly breaks the BdG symmetry. Thus the resulting Hamiltonian becomes just an Hermitian operator belonging to the 
class A. 

In the clean case, the spectrum of $\psi$-fermions remains linear in momentum $p_y$ but modes with different spin projections are split by the momentum difference $\Delta p_y{=}\mu_{\rm B} B_z/v$ (see Fig.~\ref{pic:spec1} for the energy level $\varepsilon_2$). Since in the presence of orbital magnetic field the real space coordinate is proportional to the quasimomentum in the perpendicular direction, the Zeeman field results in splitting of the chiral edge modes with different spin projections in a real space. However 
$\Delta p_y$ can be absorbed into the phase of $\psi$-fermions, thus the magnitude of $g_{\rm H}$ remains insensitive to the presence of $B_z$, see Eq. \eqref{eq:gH:sqHe}. Therefore, in the case of the Zeeman field acting at the edge only, the $2\mathbb{Z}$ quantization of $g_{\rm H}$ survives.

The edge modes have the curvature due to merging with the bulk states. Then the spectrum of spin-$\uparrow$ (spin-$\downarrow$) $\psi$-fermion floats up (down) in energy with increase of $B_z$. Hence there are energy levels (e.g. energy $\varepsilon_1$ in Fig.~\ref{pic:spec1}) for which a single edge mode remains only. Thus the spin Hall conductance becomes $g_{\rm H}{=}(2k{-}1)e^2/h$ in agreement with the $\mathbb{Z}$ quantization for the {\iqHe}.

Now let us turn on disorder at the edge again. The Zeeman 
splitting emerges 
in the {\NLSM} action as \cite{SM}
\begin{equation}\label{eq:Zedge} \mathcal{S}_{\text{e}}^{\text{(Z)}} =  i \pi  \mu_{\rm B} B_z \nu_{\text{e}} \Tr\textsf{s}_3 Q .
\end{equation}
We emphasize that the {\it physical} magnetic field $B_z$ enters the {\NLSM} as the Zeeman splitting acting in the {\it Nambu} space.
Though $\mathcal{S}_{\text{e}}^{\text{(Z)}}$ is consistent with the symmetry \eqref{eq:Qprop}, it breaks rotation symmetry in the Nambu space from SU(2) down to U(1). The term \eqref{eq:Zedge} acts as the mass term for otherwise massless theory \eqref{eq:Qedgeact}. At long distances, $|y|{\gg}1/\Delta p_y$,
the rotations $T$ commuting with the matrix $\textsf{s}_3$ survive only, enforcing diagonal form for the matrix $Q$ in the Nambu space. Substituting $Q{=}\diag\{Q_{\textsf{u}},Q_{\textsf{d}}\}$ into Eq.~\eqref{eq:Qedgeact} and using the relation $Q_{\textsf{d}}{=}{-}L_0Q_{\textsf{u}}^T L_0$, we find that $\mathcal{S}_{\rm e}$ is given by Eq.~\eqref{eq:Qedgeact} with $T$, $Q$, and $k$ substituted by $T_{\textsf{u}}$, $Q_{\textsf{u}}$, and $2k$ respectively. Since the Hermitian matrix $Q_{\textsf{u}}$ has no additional constraints except the nonlinear one, $Q_{\textsf{u}}^2{=}1$, at long distances the {\NLSM} edge action in the presence of Zeeman splitting, Eqs. \eqref{eq:Qedgeact} and \eqref{eq:Zedge}, becomes the {\iqHe} edge action with $Q_{\textsf{u}}{\in}{\rm U}(N)/[{\rm U}(N/2){\times}{\rm U}(N/2)]$. That action describes $2k$ chiral edge channels and leads to Eq. \eqref{eq:gH:sqHe} for
$g_{\rm H}$. As in the clean case, we see that within the edge theory only the Zeeman field \emph{does not change} the $2\mathbb{Z}$ quantization of $g_{\rm H}$. Thus to get the $2\mathbb{Z}{\to}\mathbb{Z}$ transformation of $g_{\rm H}$'s quantization, we have to study the bulk theory.

\noindent\textsf{\color{blue} Bulk theory for {\sqHe}.}
 Now we remind {\NLSM} describing 2D bulk of the system with the class C symmetry \cite{DellAnna,DellAnna2006,Babkin2022,Parfenov2024}
\begin{equation}
\mathcal{S}_{\text{b}} {=} {-}\frac{\bar{g}}{16}\Tr \left(\nabla Q\right)^2 {+} i\pi \bar{g}_{\rm H} \mathcal{C}
    {+}\pi \bar{\nu} \Tr [\hat{\epsilon} {+}i \mu_{\rm B} B_z \textsf{s}_3] Q .
  \label{eq:bulkacc}
\end{equation}
Here $\bar{\nu}$ denotes the bare bulk density of states and $\bar{g}$ and $\bar{g}_{\rm H}$ stand for the bare dimensional spin longitudinal and Hall conductances (in units $e^2/h$). The topology of the class C is encoded in the $\mathbb{Z}$ quantized topological charge 
\begin{equation}
     \mathcal{C}[Q] =\Tr \left(\varepsilon_{jk}Q \nabla_j Q \nabla_k Q\right)/(16\pi i), \label{eq:top:charge}
\end{equation}
where $\varepsilon_{jk}$ is the Levi-Civita symbol with $\varepsilon_{xy}{=}{-}\varepsilon_{yx}{=}1$.
For $\bar{g}_{\rm H}{=}2k$ the 
term  
 proportional to $\mathcal{C}[Q]$ in Eq. \eqref{eq:bulkacc} coincides with the first term in the edge theory \eqref{eq:Qedgeact}. 
The term in Eq. \eqref{eq:bulkacc} proportional to $B_z$ describes breaking the SU(2) symmetry in the Nambu space. 
As expected, its form is the same as for the edge theory, Eq.~\eqref{eq:Zedge} \cite{SM}.

\noindent\textsf{\color{blue} The 
crossover in the bulk.}
The {\NLSM} action \eqref{eq:bulkacc} is renormalized such that the parameters $g$, $g_{\rm H}$, and $\nu$ become length-scale ($L$) dependent. Their renormalization group (RG) equations are well-known \cite{DellAnna2006,Jeng2001,Jeng2001a,Liao2017,Babkin2022,Parfenov2024}. The class C$\to$A crossover 
can be seen already at the level of the {\NLSM} action.
At long distances, $L{\gg} L_{B}{=}\sqrt{g(L_B)/[\nu(L_B)\mu_{\rm B} B_z]}$, the Zeeman term in Eq.~\eqref{eq:bulkacc} enforces  $Q$ to become a diagonal matrix in the Nambu space. As a result, the {\NLSM} action of the class A arises. It is given by Eq. \eqref{eq:bulkacc} with $Q$, $\bar{g}$, $\bar{g}_{\rm H}$, $\bar{\nu}$ substituted by $Q_{\textsf{u}}$, $2\bar{\bar{g}}{=}2g(L_B)$, $2\bar{\bar{g}}_{\rm H}{=}2{g}_{\rm H}(L_B)$, $2\bar{\bar{\nu}}{=}2\nu(L_B)$, respectively, and with $B_z{=}0$. Thus, the {\sqHe}-to-{\iqHe} crossover can be thought roughly as follows. At $\ell{\leqslant} L{\leqslant}L_B$ the system is described by the RG equations for the class C with initial conditions $g(\ell){=}\bar{g}$ and $g_{\rm H}(\ell){=}\bar{g}_{\rm H}$. At $L{=}L_B$ the conductivities reach magnitudes, $\bar{\bar{g}}$ and $\bar{\bar{g}}_{\rm H}$, respectively. Then at $L{>}L_B$ the system is governed by the RG equations for the class A with initial conditions  $g(L_B){=}\bar{\bar{g}}$ and $g_{\rm H}(L_B){=}\bar{\bar{g}}_{\rm H}$. Consequently, a physical observable $\mathcal{O}$ at $L{>}L_B$ depends on $\bar{\bar{g}}$ and $\bar{\bar{g}}_{\rm H}$. 

The above picture of the crossover with the lengthscale $L_B$ lies in universality for some relevant in the RG sense symmetry breaking parameter (Zeeman splitting in our case \cite{GruzbergLudwig1999,BhardwajKagalovsky2015}) and, thus, applicable to both topologically trivial and topological non-trivial systems. It is easy to check that on the perturbative level the presence of $B_z$ results in the mass of diffusive modes of {\NLSM}~\eqref{eq:bulkacc} which do not belong to the class A \cite{SM}. This is exactly the mechanism that converts the perturbative part of the RG equations for the class C to the ones for the class A.  However, the topological nontrivial systems have topological excitations (instanton configurations $Q_{W}$ in our case) with integer quantized value of the topological charge $\mathcal{C}[Q_{W}]{=}W$. It is these topological excitations that are responsible for the non-perturbative part of the RG equations and for   
the periodicity of the physical observables with the bare Hall conductance in the cases of {\iqHe} \cite{pruisken1987quasiparticles,pruisken1987quasiparticlesB,pruisken1995cracking,pruisken2005instanton,pruisken2007theta} and {\sqHe} \cite{Parfenov2024}. The $2\mathbb{Z}$ ($\mathbb{Z}$) quantization in the case of {\sqHe} ({\iqHe}) implies the periodicity of physical observables with respect to $\bar{g}_{\rm H}$ ($\bar{\bar{g}}_{\rm H}$) with period $2$ ($1$). Consequently, at the {\sqHe}-to-{\iqHe} crossover the following transformation of a physical observable should occur 
\begin{equation}
\mathcal{O}
{=} \sum_{W\in \mathbb{Z}} \underbrace{\mathcal{O}^{\rm (C)}_{W} e^{i\pi \bar{g}_{\rm H} W}}_{L{\ll}L_B}  
\, 
\longrightarrow
\,
\mathcal{O} {=} 
\sum_{W\in \mathbb{Z}} \underbrace{\mathcal{O}^{\rm (A)}_{W} e^{i 2\pi  \bar{\bar{g}}_{\rm H}W}}_{L{\gg}L_B} ,
\label{eq:Puzzle}
\end{equation}
where $\mathcal{O}^{\rm (C)}_{W}{\propto} \exp({-}\pi \bar{g}|W|)$ and $\mathcal{O}^{\rm (A)}_{W} {\propto} \exp({-}2\pi \bar{\bar{g}}|W|)$.
The only consistent possibility to realize Eq. \eqref{eq:Puzzle} is the following picture of the crossover in the non-perturbative 
contributions to the RG equations.  At $L{\gg} L_B$ the non-perturbative class C contributions with odd $W$ 
have to be suppressed, while contributions with even $W$ 
transforms smoothly into the class A contributions.

\begin{figure*}[t]
\centerline{
   \includegraphics[width=.272\linewidth]{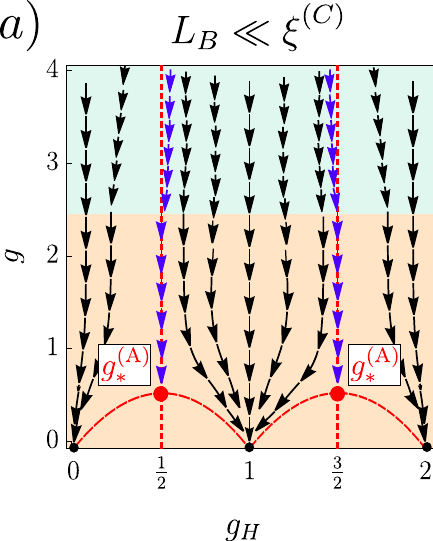}\qquad \includegraphics[width=.269\linewidth]{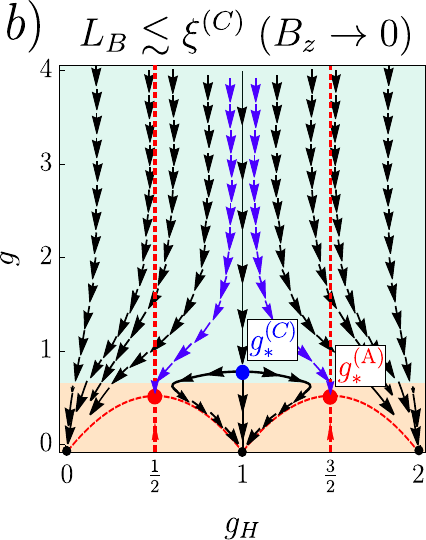}\qquad \includegraphics[width=.369\linewidth]{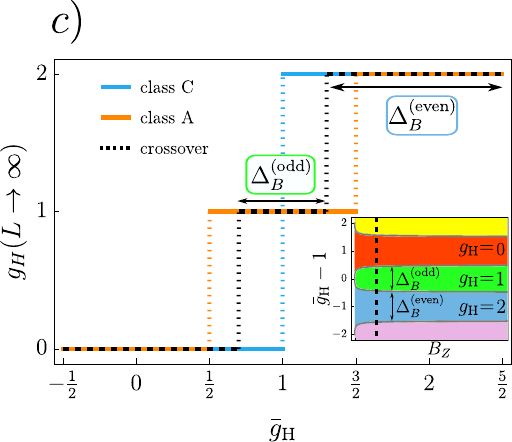}

    }
    \caption{Sketch of the crossover RG flow for strong (a) and weak (b) Zeeman splitting. For bare values $\bar{g}{\gg}1$, the RG flows in blue (orange) regions are governed by the RG equations for the class C (A). Blue lines correspond to separatrices that define the width of the {\iqHe} plateaus. (c) The quantization of the spin Hall conductance at $L{\to}\infty$ as a function of its bare value $\bar{g}_{\rm H}$ for the {\sqHe} (blue solid line), the ordinary {\iqHe} (orange solid line), and a finite $B_z$ (dashed black line). Inset shows the phase diagram in $\{\bar{g}_{\rm H},B_z\}$ plane. The black dashed line indicates the Zeeman splitting for which the main panel is plotted.}
    \label{pic:rgflow}
\end{figure*}

\noindent\textsf{\color{blue} Topological excitations at the crossover.}
To argue the above scenario, we consider the class C instantons with $W{=}1,2$. For simplicity, we present expressions for instanton solutions for $N_r{=}N_m{=}1$. The $W{=}1$ instanton is given as $Q_1{=}T^{-1} \Lambda_1(\boldsymbol{x}) T$, where $T$ represents a spatially uniform rotational matrix that defines orientation of the instanton within the NL$\sigma$M manifold and 
\cite{Parfenov2024}
\begin{equation}\label{eq:top1c}
    \Lambda_1 {=} \begin{pmatrix}
        \textsf{s}_0 \cos^2\theta  {-} \textsf{s}_1\sin^2\theta  & \frac{(i \textsf{s}_2 {-} \textsf{s}_3)}{2}e^{i \phi}\sin 2\theta \\ 
        {-}\frac{(i \textsf{s}_2 {+} \textsf{s}_3)}{2}e^{{-}i \phi}\sin 2\theta & {-}\textsf{s}_0 \cos^2\theta  {-} \textsf{s}_1\sin^2\theta 
    \end{pmatrix} .
\end{equation}
Here $\theta{=}\arctan(\lambda/|z{-}z_0|)$, $\phi{=}\arg(z{-}z_0)$, $z{=}x{+}i y$ is the complex coordinate, $\lambda$ is the instanton scale size,  and $z_0$ is the position of instanton. In the absence of the last term in 
Eq.~\eqref{eq:bulkacc}, we find $\mathcal{S}_{\text{b}}[Q_1]{=}{-}\pi \bar{g} {+} i \pi \bar{g}_{\rm H}$, such that the parameters $\lambda$, $z_0$, and the generators of the $T$-rotations constitute the zero mode manifold of the $W{=}1$ instanton.   
Due to the Zeeman term, 
rotational zero modes with $[T,\textsf{s}_3]{\ne} 0$ acquire a mass ${\propto}B_z \ln(L/\lambda)$, i.e. modification of the zero mode manifold from $T{\in}{\rm Sp}(2N)$ (class C) to $T{\in}{\rm U}(N)/[{\rm U}(N/2){\times}{\rm U}(N/2)]$ (class A) occurs. However, $\lambda$ and $z_0$ remain zero modes, i.e. $W{=}1$ instanton is not fully suppressed by $B_z$ at the classical level.

Accounting for fluctuations around the $W{=}1$ instanton leads to logarithmically divergent renormalizations in physical observables. These divergences can be resummed within the RG framework. Without the Zeeman splitting, the resummation process continues until the RG flow reaches a scale where the instanton size becomes comparable to the dynamically generated localization length in the class C, $\lambda {\sim} \xi^{(C)}{\simeq}\ell \exp(\pi\bar{g})$ \cite{Parfenov2024}. 
In the presence of a non-zero $B_z$, the RG procedure for $W{=}1$ instantons halts at $\lambda{\sim}L_B$, because all instantons with sizes $\lambda{>}L_B$ fail to contribute logarithmic corrections to physical observables \cite{SM}. It leads to suppression of contributions from the $W{=}1$ instantons to the RG equations beyond the lengthscale $L_B$.

The $W{=}2$ instanton solution can be expressed as $Q_2 {=} \tilde{T}^{-1}\Lambda_2(\boldsymbol{x}) \tilde{T}$, where the matrix $\tilde{T}$ contains rotational zero modes and (for $N_r{=}N_m{=}1$ as above)
    \begin{gather}
        \Lambda_2(\boldsymbol{x}) = \begin{pmatrix}
         \hat{1} & 0 \\
        0 & -\hat{1}
    \end{pmatrix}\mathcal{K}^{-1} \begin{pmatrix}
        \cos 2\hat{\theta} & \sin 2\hat{\theta} \\
        -\sin 2\hat{\theta} & \cos 2\hat{\theta}
    \end{pmatrix}\mathcal{K} .  \label{eqn:06:gsEven}
    \end{gather}
Here $\mathcal{K}{=}\text{diag}\{\mathcal{U},\mathcal{U}^*\}$ with an arbitrary U(2) matrix $\mathcal{U}$
and a 
matrix $\hat{\theta} {=}\text{diag}\{\theta_1,\theta_2\}$.  The instanton angles $\theta_{j}$ are defined similarly to those of the $W{=}1$ instanton and involve two sets of zero modes $z_{0}^{(j)}$ and $\lambda_j$. The resulting classical bulk action for this solution can be divided into two parts: the classical action for 
$W{=}2$ instanton, $\mathcal{S}_{\rm b}^{\text{(cl)}}{=}{-}2\pi \bar{g}{+}2 \pi i\bar{g}_{\rm H}$, and the Zeeman contribution, $\mathcal{S}_{\rm b}^{\rm (Z)} {\propto} B_z \int d^2\bm{x}  [\cos2\theta_1(\boldsymbol{x}) {-}\cos2\theta_2(\boldsymbol{x})]$, with a coefficient depending on the matrices $\tilde{T}$ and 
$\mathcal{U}$ (see End Matter and \cite{SM} for details). The Zeeman term forces the synchronization of instanton scale sizes, 
$\lambda_1{=}\lambda_2$, and positions, $z_{0}^{(1)}{=}z_{0}^{(2)}$. 
Then the $W{=}2$ instanton becomes diagonal matrix in the Nambu space,  
which can be interpreted as the two independent class A $W{=}1$ instantons. Thus 
the Zeeman splitting forces 
the class C $W{=}2$ instanton to transform into the class A $W{=}1$ instanton already at the level of the classical action (see End Matter). 

\color{black}

\noindent\textsf{\color{blue} Physical predictions.} The above picture of the {\sqHe}-to-{\iqHe} crossover 
has implications for the length-scale dependence of physical observables. For example,
the dependence of $g$ and $g_{\rm H}$ on $L$ 
can be visualized as a two-parameter scaling diagram shown in Fig.~\ref{pic:rgflow}. We assume that RG flow starts from a weak coupling region, $\bar{g}{\gg}1$. At strong Zeeman splitting, $L_B{\ll}\xi^{(C)}$, \footnote{We note that the strongest magnetic field $B_z$ we consider corresponds to the condition $L_B {\sim}\ell$. For stronger fields the crossover occurs at the ballistic length scales. Also we assume that the Zeeman field corresponding to $L_B {\sim}\ell$ is not enough to destroy the superconducting state.} the crossover 
occurs in weak coupling region, $\bar{\bar{g}}{\gg}1$, see Fig.~\ref{pic:rgflow}a. In contrast, at weak $B_z$ such that $L_B{\lesssim}\xi^{(C)}$, the crossover occurs in the strong coupling regime close to the class C unstable fixed point $g_*^{(C)}{=}\sqrt{3}/2$ \cite{Evers1997,Cardy2000} (
for class A
$g_*^{(A)}{\simeq}0.5{\div}0.6$ \cite{Huo1993,Gammel1994,Schweitzer2005}), see Fig.~\ref{pic:rgflow}b. In both cases of strong and weak $B_z$, the flow lines starting at $|\bar{g}_{\rm H}{-}1|{\leqslant}\Delta^{\rm (odd)}_B/2$ approach the stable fixed point at $g{=}0$ and $g_{\rm{H}}{=}1$ as $L{\to}\infty$.
Thus for $B_z{\neq}0$ the RG flow in Fig.~\ref{pic:rgflow} shows the $\mathbb{Z}$ quantization of $g_{\rm{H}}$
as $L{\to}\infty$. 
The RG flow in Fig.~\ref{pic:rgflow} looks similar to the crossover RG flow due to breaking of spin degeneracy in an ordinary {\iqHe} \cite{PruiskenSpin}
and mixing of valleys for the {\iqHe} in graphene \cite{Ostrovsky2008}. However, those crossovers occur within the same class A.

For $|\bar{g}_{\rm H}{-}1|{\leqslant}\Delta^{\rm (odd)}_B/2$ the dependence of $g_{\rm H}$ on $L$ is non-monotonous with the extremum at $L{\sim}L_B$. Plateaus at odd integer values in dependence of $g_{\rm H}$ on $\bar{g}_{\rm H}$ start to develop as $L$ grows beyond $L_B$. In the limit $L{\to}\infty$  
the dependence of $g_{\rm H}$ on $\bar{g}_{\rm H}$ becomes a step-like with plateaus at $\mathbb{Z}$, see Fig.~\ref{pic:rgflow}c. However, the widths of the odd, $\Delta^{\rm (odd)}_B$,  and even, $\Delta^{\rm (even)}_B{=}2{-}\Delta^{\rm (odd)}_B$, plateaus are different. This fact reflects periodic dependence of physical observables on $\bar{g}_{\rm H}$ with period $2e^2/h$ as follows from Eq.~\eqref{eq:Puzzle}. 

At  $B_z{\to}0$, the width of the odd plateaus can be estimated as $\Delta^{\rm (odd)}_B{\sim} |B_z|^{3/7}$ \cite{Gruzberg1999}. 
At strong Zeeman splitting, $L_B{\ll}\xi^{(C)}$, the odd-plateau width approach $1$ as  $1{-}\Delta^{\rm (odd)}_B{\sim} [(L_B{-}\ell)/\xi^{(C)}]\ln^3 (\xi^{(C)}/L_B)$  \cite{SM}. 
We note that the staircase with
$\Delta^{\rm (odd)}_B{\neq}\Delta^{\rm (even)}_B$
distinguishes the {\iqHe} obtained in a result of crossover from the {\sqHe} in a topological superconductor, and the ordinary {\iqHe}. 

\color{black}

\noindent\textsf{\color{blue} Summary.} We developed the coherent physical picture of  the spin-to-integer quantum Hall effect crossover in the bulk and at the edge of the topological superconductors. 
We demonstrated that it is not possible to describe the crossover in terms of the edge theory only. The correct description of the crossover involves the bulk theory, in particular, topological excitations (instantons). We found that although the spin Hall conductance becomes quantized in units $e^2/h$ as a result of the crossover, the periodic dependence of the  physical observables on the bare spin Hall conductance has the period $2e^2/h$ as for the {\sqHe}. We obtained that after the crossover the widths of the odd and even {\iqHe} plateaus are different in contrast to the conventional {\iqHe} staircase.    
Although we study {\sqHe}-to-{\iqHe} crossover in the absence of electron-electron interaction, we expect that it does not alter the developed physical picture.

Finally, we mention that
twisted Bi$_2$Sr$_2$CaCu$_2$O$_{8{+}x}$ bilayers have been recently shown to spontaneously break time-reversal symmetry \cite{Zhao-Science}, in agreement with theoretical predictions for emergent $d_{x^2{-}y^2}{+}i d_{xy}$ topological superconducting state \cite{Franz,Volkov}. Results of our work suggest that a magnetic field parallel to bilayers is an efficient tool to control and manipulate the edge spin-current-carrying states in such topological superconductors in a way similar to manipulation of edge current channels in the conventional {\iqHe} \cite{Altimiras2010,Altimiras20102,Heiblum,Carrega}.

\noindent\textsf{\color{blue}Acknowledgements.} We thank D. Antonenko, I. Gornyi, I. Gruzberg, A. Mirlin, and P. Ostrovsky for useful
discussions and comments. The work was funded in part by Russian Ministry of Science
and Higher Education (Project No. FFWR-2024-0017) as well
as by Basic research program of HSE. The authors acknowledge the hospitality during the “Nor-Amberd School in Theoretical Physics 2024” where part of this work has been performed. M.V.P.
and I.S.B. acknowledge personal support from the Foundation
for the Advancement of Theoretical Physics and Mathematics
“BASIS”. I.S.B. is grateful to Institute for Theoretical and Mathematical Physics MSU for hospitality.

\bibliography{literature_classC_top}

\onecolumngrid
    \begin{center} 
    {\bf \large End Matter}
\end{center}

\twocolumngrid
\appendix*

\setcounter{equation}{0}
\renewcommand{\theequation}{A\arabic{equation}}
\noindent\textsf{\color{blue} Appendix A. The $W{=}2$ instanton and the Zeeman term.}
The $W{=}2$ instanton can be expressed as follows $Q_2(\boldsymbol{x}){=}\tilde{T}^{-1}\Lambda_2(\bm{x})\tilde{T}$, where $\Lambda_2(\bm{x})$ is given by Eq. \eqref{eqn:06:gsEven}.  
We stress that there exist two independent sets of scale sizes $\lambda_{1,2}$ and instanton positions $z^{(1,2)}_0$. This can be interpreted as two independent instanton solutions with mixing in Nambu space, introduced by the matrix $\mathcal{K}$. We decompose the spatially uniform rotation $ \tilde{T}$ into two parts: $ \tilde{T}= U^{-1} T$, where $U$ is a non-unitary rotation derived in \cite{Parfenov2024} to simplify the solution of the self-duality (Belavin-Polyakov) equation:
\begin{equation}
 \begin{split}
     \nabla_x Q_2 \pm i Q_2\nabla_y Q_2  &=0,  \; 
     U  =  \frac{1}{\sqrt{2}}\begin{pmatrix}
        1 & 0 & 1 & 0 \\
        1 & 0 & -1 & 0 \\
        0 & -1 & 0 & 1 \\
        0 & 1 & 0 & 1
        \end{pmatrix} 
 \end{split}   
 \label{eq:em:RGcrossover1}
\end{equation}
and $T$ is the unitary component, representing the rotational zero modes.

The exact parametrization of the U(2) matrix $\mathcal{U}$ (involved in the definition of $\mathcal{K}$), which satisfies the self-duality equation, is given by:
\begin{equation}
     \mathcal{U} = e^{i \varphi/2}\begin{pmatrix}
        e^{i\alpha} \cos \chi &  e^{i\beta} \sin \chi \\
        -e^{-i\beta} \sin \chi & e^{-i\alpha} \cos \chi
        \end{pmatrix} ,
\end{equation}
where  $\alpha {=} \beta {=}\delta_-(z)$ and $\varphi {=}2\delta_+(z)$ are spatially dependent angles with $\delta_{\pm} {=}[\arg(z{-}z^{(1)}_0){\pm}\arg(z{-}z^{(2)}_0)]/4$.
We note that the mixing angle $\chi$ is a zero mode of this solution, along with $z_0^{(i)}$ and $\lambda_i$. Substituting this solution into the symmetry-breaking part of action $S_B$, we obtain:
\begin{multline}
\mathcal{S}_{\rm b}^{\rm (Z)} = -4i \pi \bar{\nu} \mu_B B_z \sin2\chi \\ \times \int  \frac{(\lambda_1^2|z-z_0^{(2)}|^2- \lambda_2^2|z-z_0^{(1)}|^2)dx dy}{(|z-z_0^{(1)}|^2+\lambda_1^2)(|z-z_0^{(2)}|^2+\lambda_2^2)} .
\end{multline}
This integral exhibits a logarithmic divergence at large scales. Using a 
regularization scheme with infra-red length scale being the system size $L$, we compute the integral and obtain
\begin{gather}
   \mathcal{S}_{\rm b}^{\rm (Z)} = -8i \pi^2 \bar{\nu} \mu_B B_z \sin2\chi \Bigl[\lambda_1^2 \ln\frac{L}{\lambda_1} -\lambda_2^2 \ln\frac{L}{\lambda_2} \notag \\-\frac{1}{2 L^2} \left(\lambda_2^4-\lambda_1^4+\lambda_1^2 |z^{(1)}_0{-}z^{(2)}_0|^2\right)\Bigr ] ,
\end{gather}
We take into account the leading non-logarithmic correction to demonstrate that both the instanton positions and instanton scales cease to be zero modes. We point two possibilities for disappearing of $\mathcal{S}_{\rm b}^{\rm (Z)}$: synchronization of instanton scale sizes, $\lambda_1 {=} \lambda_2$, and instanton positions, $z^{(1)}_0{=}z^{(2)}_0$, or
vanishing the factor 
$\sin 2\chi$. In order to resolve this we employ variational principle that indictates 
that the former is more preferable, i.e.  $\lambda_1 {=} \lambda_2$ and $z^{(1)}_0{=}z^{(2)}_0$ (for details see \cite{SM}).

Then the two-instanton solution reduces to two independent $W{=}1$ instantons of the class A: 
\begin{equation}
      \tilde{\Lambda}_2(\boldsymbol{x}) =\begin{pmatrix}
       \Lambda^{(A)}_1(\boldsymbol{x},\lambda,z_0) & 0  \\ 
        0 & \Lambda^{(A)}_1(\boldsymbol{x},\lambda,z_0) 
    \end{pmatrix},
    \label{eq: inst1}
\end{equation}
with $\lambda{=}\lambda_1{=}\lambda_2$. Here we define 
the instanton for class A, $\Lambda^{(A)}_1(\boldsymbol{x},\lambda,z_0)$, in the following form:
\begin{equation}
    \Lambda^{(A)}_1(\boldsymbol{x},\lambda,z_0) = \begin{pmatrix}
        \cos 2\theta  & e^{i \phi}\sin 2\theta \\ 
        e^{-i \phi}\sin 2\theta & -\cos 2\theta
    \end{pmatrix} ,
\end{equation}
where $\theta{=}\arctan{\lambda/(z{-}z_0)}$ and $\phi{=}\arg(z{-}z_0)$.

\setcounter{equation}{0}
\renewcommand{\theequation}{B\arabic{equation}}
\noindent\textsf{\color{blue} Appendix B. The crossover RG flow and the plateaus' widths.}  In the weak coupling 
regime, $\bar{g} \gg 1$, we analyze the 
crossover using the 
RG framework. The corresponding RG equations for spin conductivities can be combined from the known non-perturbative RG equations for the class C and the class A:
\begin{equation}
 \begin{split}
     \frac{dg}{d\ln L}  & =-\beta_{g}^{\text{(C)}}(1-f_X)-\beta_{g}^{\text{(A)}} f_X , \\
     \frac{dg_{\rm H}}{d\ln L}  & = -\beta_{g_{\rm H}}^{\text{(C)}}(1-f_X)-\beta_{g_{\rm{H}}}^{\text{(A)}} f_X .
 \end{split}   
 \label{eq:em:RGcrossover1}
\end{equation}
Here the beta-functions for the classes C and A are given as \cite{Hikami1981,pruisken2005instanton,Parfenov2024}
\begin{equation}
\begin{split}
\beta_{g}^{\text{(C)}}&{\simeq}\frac{1}{\pi}{+}\mathcal{D}_{\text{C}}(\pi g)^3 e^{{-}\pi g}\cos \pi  g_{\rm H},\\ 
\beta_{g_{\rm H}}^{\text{(C)}}&{\simeq}\mathcal{D}_{\text{C}}(\pi g)^3 e^{{-}\pi g}\sin \pi  g_{\rm H},\\    \beta_{g}^{\text{(A)}}&{\simeq}\frac{1}{2\pi^2 g}{+}\mathcal{D}_{\text{A}}(\pi g)^2 e^{{-}2\pi g}\cos 2 \pi  g_{\rm H},\\ 
\beta_{g_{\rm H}}^{\text{(A)}}&{\simeq}\mathcal{D}_{\text{A}}(\pi g)^2 e^{{-}2\pi g}\sin 2 \pi  g_{\rm H}
\end{split}
\end{equation}
Here $\mathcal{D}_{\text{C}}{=}8e^{{-}2{-}\gamma}/\pi$ and $\mathcal{D}_{\text{A}}{=}4\pi/e$ are  
numerical constants derived within the Pauli-Villars regularization scheme~\footnote{In Ref. \cite{pruisken2005instanton} the 4 times larger result for $\mathcal{D}_{\text{A}}$ has been reported erroneously.}. 
In comparison with perturbative weak-localization corrections the instanton corrections in  
$\beta_{g}^{\text{(C,A)}}$ 
can be neglected. 
The sharpness of the crossover transition allows us to 
approximate the crossover function as Heaviside step-function: $f_X {=} \theta(L{-}L_B)$ (for a detailed explanation, see \cite{SM}). The RG flow corresponding to these equations is presented in Fig.\ref{pic:rgflow} 
in the main text. Solutions for Eqs. \eqref{eq:em:RGcrossover1} is presented in \cite{SM}. 

The widths of the plateaus can be estimated by analyzing the RG flow's critical trajectories (depicted as blue lines in Fig.\ref{pic:rgflow}a and Fig.\ref{pic:rgflow}b). The starting point of a critical trajectory corresponds to the minimal deviation in the bare value of $\bar{g}_{\rm H}$ from $1$ that leads to a departure from class C behavior in the infrared limit ($L{\rightarrow} \infty$). By solving Eq.~\eqref{eq:em:RGcrossover1}, we obtain this deviation \cite{SM}: 
\begin{equation}
   \bar{g}_{\rm H}  = \frac{2}{\pi} \tan^{-1} \left(e^{-\pi \mathcal{D}_{\text{C}} f(\bar{\bar{g}},\bar{g})}\right).
\end{equation}
The limiting cases of this expression are discussed in the main text.


\newpage



\section*{Supplementary material}

# ONLINE SUPPORTING INFORMATION
# Bulk-edge correspondence at the spin-to-integer quantum Hall effect crossover

M. V. Parfenov[1,2] and I. S. Burmistrov[1,2]

[1] *L. D. Landau Institute for Theoretical Physics, Semenova 1-a, 142432, Chernogolovka, Russia*
[2] *Laboratory for Condensed Matter Physics, HSE University, 101000 Moscow, Russia*

In this notes we present details of (i) derivation of the NL$\sigma$M at the edge, (ii) spectrum of the toy model, (iii) suppression of the diffusive modes by the Zeeman splitting in the perturbation theory, (iv) suppression of the $W=1$ class C instanton due to the quantum fluctuations, (v) variational anzats for the $W=2$ instanton, and (vi) estimates for the plateau width.

## S.I. DERIVATION OF THE EDGE NL$\sigma$M

In this section, we demonstrate the derivation of the NL$\sigma$M for quasiparticles at the edge of $d_{x^2-y^2}+id_{xy}$ superconductor. We begin with the action eq. (1) in the main text. Using a bivector representation (with the aid of an additional Nambu-type space), this action can be rewritten in a more transparent form:

$$S = \int_0^\beta d\tau \int dy \overline{\Psi} \left(-\partial_\tau - H_0 - V\right) \Psi, \quad \Psi = \frac{1}{\sqrt{2}} \begin{pmatrix} \psi_\uparrow \\ \psi_\downarrow \\ \overline{\psi}_\uparrow^\dagger \\ \overline{\psi}_\downarrow^\dagger \end{pmatrix}, \tag{S.1}$$

in this equation Hamiltonian has the form (in $4\times 4$ matrix space):

$$H = \begin{pmatrix} v\hat{p}_y + \eta_3 & 0 & 0 & \eta_1 - i\eta_2 \\ 0 & v\hat{p}_y + \eta_3 & -\eta_1 + i\eta_2 & 0 \\ 0 & -\eta_1 - i\eta_2 & v\hat{p}_y - \eta_3 & 0 \\ \eta_1 + i\eta_2 & 0 & 0 & v\hat{p}_y - \eta_3 \end{pmatrix}. \tag{S.2}$$

This Hamiltonian has both SU(2) and particle-hole symmetries: $\mathsf{s}_1 (H)^T \mathsf{s}_1 = -H$, $\sigma_2 H \sigma_2 = H$. Using a representation in terms of spin fermions $\chi$, we can rotate the Hamiltonian in Nambu space to simplify its structure. Let us write:

$$\begin{pmatrix} \overline{\chi} & \chi^T \end{pmatrix} \begin{pmatrix} H & 0 \\ 0 & \sigma_2 H \sigma_2 \end{pmatrix} \begin{pmatrix} \chi \\ \overline{\chi}^T \end{pmatrix} = \begin{pmatrix} \overline{\chi} & \chi^T \sigma_2 \end{pmatrix} H \mathsf{s}_0 \begin{pmatrix} \chi \\ \sigma_2 \overline{\chi}^T \end{pmatrix},$$
$$\Phi = \frac{1}{\sqrt{2}} \begin{pmatrix} \chi \\ \sigma_2 \overline{\chi}^T \end{pmatrix}, \quad H = -iv\sigma_0 \partial_y + \boldsymbol{\eta}\boldsymbol{\sigma}, \tag{S.3}$$

here $\mathsf{s}_i$ is Pauli matrices in Nambu-type space. This Hamiltonian exhibits a standart class C symmetry property: $H = -\sigma_2 H^T \sigma_2$. One can check charge conjugation relation for bivector $\Phi$:

$$\overline{\Phi} = [C\Phi]^T, \quad C = \sigma_2 \otimes (i\mathsf{s}_2), \quad C^2 = -1. \tag{S.4}$$

Therefore, resulting action has the form:

$$S = \int_0^\beta d\tau \int dy \overline{\Phi} \left(-\partial_\tau \mathsf{s}_0 - H \mathsf{s}_0\right) \Phi. \tag{S.5}$$

After that we should introduce Matsubara space for fermionic fields $\psi$ with frequences $\varepsilon_n = \pi T (2n+1)$:

$$\psi_i = \sum_m \psi_{m,i} e^{-i\varepsilon_m \tau}, \quad \overline{\psi}_i = \sum_n \overline{\psi}_{n,i} e^{+i\varepsilon_n \tau}. \tag{S.6}$$

In terms of $\Phi$ bivector:

$$\overline{\Phi}^\alpha \to \sum_n \overline{\Phi}_n^\alpha e^{i\sigma_3 \varepsilon_n \tau} \mathsf{s}_0, \quad \Phi^\alpha \to \sum_m e^{-i\sigma_3 \varepsilon_m \tau} \mathsf{s}_0 \Phi_m^\alpha. \tag{S.7}$$

After integrating over imaginary time for the part of the action that commutes with $\mathsf{s}_3$, we obtain:

$$S^{(0)} = \beta \int dy \sum_n \overline{\Phi}_n^\alpha \left(i\varepsilon_n \mathsf{s}_0 \sigma_3 + iv\sigma_0 \mathsf{s}_0 \partial_y\right) \Phi_n^\alpha. \tag{S.8}$$

One can observe the emergence of a nontrivial structure in spin space for the energy term. Due to the noncommutative nature of the random potential contribution, integration over imaginary time leads to a nontrivial structure in Matsubara space:

$$S^{(\mathrm{dis})} = -\beta \int dy \sum_{n,m} \overline{\Phi}_n^\alpha \mathsf{s}_0 \left(\eta_1 \sigma_1 + \eta_2 \sigma_2\right) \delta_{m,-n-1} \Phi_m^\alpha - \beta \int dy \sum_n \overline{\Phi}_n^\alpha \mathsf{s}_0 \eta_3 \sigma_3 \Phi_n^\alpha. \tag{S.9}$$

For simplicity, we define new matrix in Matsubara space $l_0 = \delta_{m,-n-1} = \delta_{\varepsilon_m, -\varepsilon_n}$. To simplify further computations, we perform a matrix rotation in the combined Matsubara and spin spaces:

$$\Xi_m^\alpha = U_\xi \Phi_m^\alpha, \quad U_\xi = \begin{pmatrix} 1 & 0 \\ 0 & l_0 \end{pmatrix}_\mathsf{s},$$
$$\overline{\Xi} = \Xi^T \mathcal{C}^T, \quad \mathcal{C} = (i\mathsf{s}_2) l_0 \sigma_2 = \mathcal{C}^T. \tag{S.10}$$

After this rotation, we rewrite each part of the action in terms of the fermion fields $\Xi$, defined above:

$$S^{(0)} = \beta \int dy \sum_n \overline{\Xi}_n^\alpha \left(i\varepsilon_n \sigma_0 \mathsf{s}_0 + iv\mathsf{s}_0 \sigma_0 \partial_y\right) \Xi_n^\alpha, \tag{S.11}$$

$$S^{(\mathrm{dis})} = -\beta \int dy \sum_n \overline{\Xi}_n^\alpha \sigma_0 \left(\eta_1 \sigma_1 + \eta_2 \sigma_2 + \eta_3 \sigma_3\right) \Xi_n^\alpha = -\beta \int dy \sum_n \overline{\Xi}_n^\alpha \mathsf{s}_0 V \Xi_n^\alpha, \tag{S.12}$$

where correlator of random potential has the following form:

$$\langle V_{\alpha\beta}(y) V_{\gamma\delta}(y') \rangle = \varkappa \left(2\delta_{\alpha\delta}\delta_{\beta\gamma} - \delta_{\alpha\beta}\delta_{\gamma\delta}\right), \tag{S.13}$$

using Wick theorem we average disorder part of action over random potential:

$$\left\langle \frac{1}{2} \left(S^{(\mathrm{dis})}\right)^2 \right\rangle = \frac{\varkappa \beta^2}{2} \int dy \left(2 \overline{\Xi}_\alpha^j \Xi_\beta^j \overline{\Xi}_\beta^{j'} \Xi_\alpha^{j'} - \overline{\Xi}_\alpha^j \Xi_\alpha^j \overline{\Xi}_\beta^{j'} \Xi_\beta^{j'}\right), \tag{S.14}$$

where $j$ and $j'$ are collective indices for Nambu, Matsubara, and replica spaces, while $\alpha$ and $\alpha'$ denote spin indices. The second term in the brackets vanishes by definition of $\Xi$. It is convenient to rewrite the above expression in terms of the trace over Nambu, Matsubara, and replica spaces:

$$\left\langle S^{(\mathrm{dis})} \right\rangle = -\varkappa \beta^2 \int dy \mathrm{Tr}\left[\Xi_\beta \overline{\Xi}_\beta \Xi_\alpha \overline{\Xi}_\alpha\right]. \tag{S.15}$$

As a consequence of spin invariance of the initial Hamiltonian, $\Xi_\beta \overline{\Xi}_\beta$ becomes trivial in spin space. The next step is to decouple four-fermion term using the Hubbard-Stratonovich transformation. Introducing hermitian matrix field $\mathcal{Q}$, we obtain:

$$S_Q = -\frac{1}{4\varkappa} \int dy \mathrm{Tr} \mathcal{Q}^2 + i\beta \int dy \overline{\Xi} \mathcal{Q} \sigma_0 \Xi. \tag{S.16}$$

From this form of action and eq. (S.9) we can obtain BdG symmetry for $Q$-field:

$$\mathcal{Q} = -\mathsf{s}_2 l_0 \mathcal{Q}^T l_0 \mathsf{s}_2. \tag{S.17}$$

After integration over Grassmanian fields $\Xi$, we obtain resulting action for matrix field $Q$:

$$S_Q = -\frac{1}{4\varkappa} \int dy \, \mathrm{Tr} \, \mathcal{Q}^2 + \int dy \, \mathrm{Tr} \log\left[i\mathsf{s}_0 \hat{\varepsilon}_n + iv\mathsf{s}_0 \mathbf{1}_M \partial_y + i\mathcal{Q}\right]. \tag{S.18}$$

In the absence of symmetry-breaking terms and at zero temperature ($\varepsilon_n \to 0$) we can write down the saddle-point equation:

$$-\frac{1}{2\varkappa}\mathcal{Q} + i\left[iv\partial_y + i\mathcal{Q}\right]^{-1} = 0. \tag{S.19}$$

The solution of this equation takes the following form:

$$(\mathcal{Q}_{\rm sp})^{\alpha\beta} = \frac{1}{2\tau}\Lambda^{\alpha\beta}, \quad \Lambda^{\alpha\beta} = \tau_3 s_0 \delta^{\alpha\beta}, \tag{S.20}$$

where $\tau_3$ is the third Pauli matrix in retarded-advanced space and $\tau$ denotes mean free time, which can be self-consistently expressed in terms of the random potential correlator and edge velocity: $1/\tau = 2\varkappa/v$.

To obtain the diffusive NL$\sigma$M for edge quasiparticles, we expand around the metallic saddle point. For this purpose, we parametrize the sigma-model manifold in terms of non-uniform unitary rotations around the saddle point:

$$\mathcal{Q} = \frac{1}{2\tau}T^{-1}\Lambda T, \tag{S.21}$$

where $T$ are massless modes, which define target manifold of class C: Sp(2n)/U(n). Logarithmic part of action, therefore, can be rewritten in the following form:

$$S_Q^{\log} = \mathrm{Tr}\log\left[\mathcal{B}[T] + G_0^{-1}\right], \quad G_0^{-1} = -H_0 + i\Lambda/2\tau, \quad \mathcal{B}[T] = ivT\partial_y T^{-1} + i\epsilon T\tau_3 T^{-1}, \tag{S.22}$$

for simplicity here we switch from Matsubara representation to retarded-advanced. Making a gradient expansion in $T$ we can obtain:

$$S_Q^{\log} \approx \mathrm{Tr}\,\mathcal{B}[T]G_0 - \frac{1}{2}\mathrm{Tr}\,\mathcal{B}[T]G_0\mathcal{B}[T]G_0. \tag{S.23}$$

The first term gives rise to a topological term in its edge form, along with a source term:

$$S_{\rm top} = \frac{1}{2}\int dy\,\mathrm{Tr}\left(T\partial_y T^{-1}\Lambda\right), \quad S_\Lambda = \frac{\epsilon}{2v}\int dy\,\mathrm{Tr}\,\Lambda Q, \tag{S.24}$$

here we use definition of Green function for edge fermions:

$$G_0(y,y') = \int\frac{dp}{2\pi}\frac{e^{ip(y-y')}}{-vp + i\Lambda/2\tau} = -\frac{ie^{-|y-y'|/2v\tau}}{2v}\left(\mathrm{sgn}(y-y') + \Lambda\right), \tag{S.25}$$

and here we define edge $Q$-matrix: $Q = T^{-1}\Lambda T$ with constraint $Q^2 = 1$. Second order term leads to kinetic part of NL$\sigma$M action:

$$\frac{v^2}{2}\int_{y,y'}\frac{dpdp'}{(2\pi)^2}\mathrm{Tr}\left[U[T]\frac{e^{ip(y-y')}}{-vp + i\Lambda/2\tau}U[T]\frac{e^{-ip'(y-y')}}{-vp' + i\Lambda/2\tau}\right], \tag{S.26}$$

here $U[t] = T\partial_y T^{-1}$. In diffusive regime elastic scattering length $l = v\tau$ is small, therefore we can put both $U[T]$ at the same point. After that we can integrate this expression and obtain:

$$S_q = -\frac{l}{8}\int dy\,\mathrm{Tr}\,(\partial_y Q)^2 = -\frac{v^2}{16\varkappa}\int dy\,\mathrm{Tr}\,(\partial_y Q)^2. \tag{S.27}$$

Resulting expression for action at the edge can be written in the following form:

$$\mathcal{S}_e = -\frac{v^2}{16\varkappa}\int dy\,\mathrm{Tr}\,(\partial_y Q)^2 + \frac{1}{2}\int dy\,\mathrm{Tr}\left(T\partial_y T^{-1}\Lambda\right) + \frac{\epsilon}{2v}\int dy\,\mathrm{Tr}\,\Lambda Q. \tag{S.28}$$

Introducing of Zeeman field can be done in the following way:

$$S_e^{(Z)} = \mu_B B_z \sum_{\alpha=1}^{N_r}\int_0^\beta d\tau \int dy\,\overline{\psi}^\alpha \sigma_3 \psi^\alpha. \tag{S.29}$$

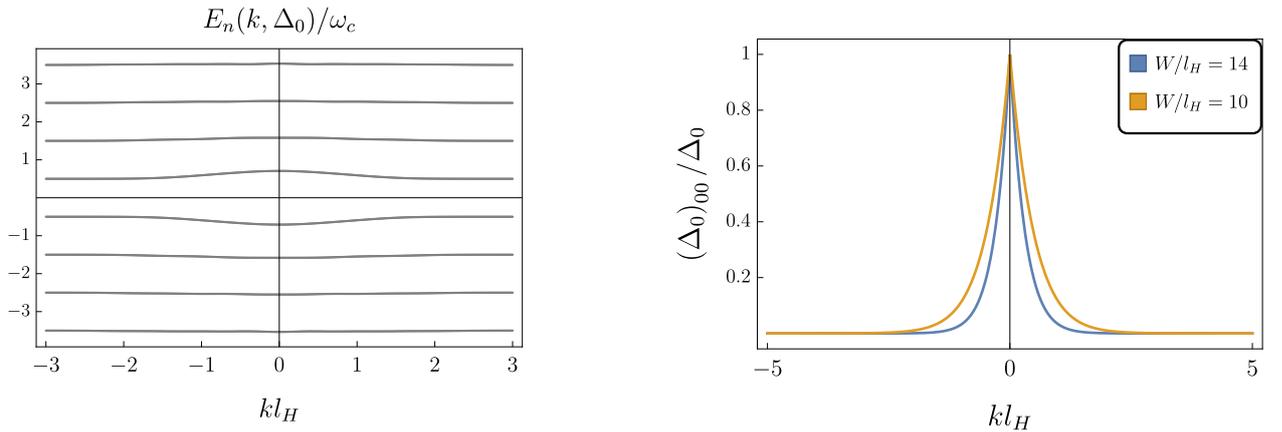

FIG. 1. Left panel: Spectrum of Landau levels in the presence of a constant pairing potential $\Delta_0$ in an infinite system. A nonzero $\Delta_0$ induces a dispersion of the Landau bands. Right panel: Diagonal matrix element of the pairing potential at the lowest Landau level within a stripe of width $W$.

After all basis changes, described above, one can write this term in $\Xi$ fermion representation:

$$S_e^{(Z)} = -\beta\mu_B B_z \int dy \bar{\Xi}_n^\alpha \mathsf{s}_3 \Xi_n^\alpha. \tag{S.30}$$

Integration over Grassmanian fields leads to appearance of term under Tr log:

$$S_Q^{\log} = \int dy \, \mathrm{Tr}\log\left[i\mathsf{s}_0\hat{\varepsilon}_n + iv\mathsf{s}_0\mathbf{1}_M \partial_y - \mu_B B_z \mathsf{s}_3 + i\mathcal{Q}\right]. \tag{S.31}$$

By following the same steps as for the density of state contribution and using equation (S.23), we obtain:

$$\mathcal{S}_e^{(Z)} = \frac{i\mu_B B_z}{2v} \int dy \, \mathrm{Tr}\, \mathsf{s}_3 Q. \tag{S.32}$$

### S.II. LANDAU LEVELS FOR UNIFORM PAIRING $\Delta_0$

We consider a toy model for the Bogoliubov–de Gennes (BdG) Hamiltonian describing quasiparticles in a sufficiently strong magnetic field with a nonzero induced order parameter $\Delta$. For simplicity, we assume a uniform amplitude of the order parameter. In a real system, however, the self-consistency equation for $\Delta$ must be taken into account [1]. Under this approximation, the single-particle BdG Hamiltonian takes the form:

$$\mathcal{H} = \begin{pmatrix} H_0 & \Delta_0 \\ \Delta_0 & -H_0^T \end{pmatrix}, \quad H_0 = \frac{1}{2m}\left(-i\boldsymbol{\nabla} - e\boldsymbol{A}\right)^2. \tag{S.33}$$

We first analyze the case of an infinite system. In the Landau gauge ($A_x = 0$, $A_y = -H_z x$), the electron eigenfunctions take the well-known form:

$$\psi_{n,k}(x,y) = \frac{1}{\sqrt{L}} e^{iky} \phi_n(x - x_k), \quad \phi_n(x) = \left(\frac{1}{2^n n! l_H \sqrt{\pi}}\right)^{1/2} \exp\left(-\frac{x^2}{2l_H^2}\right) H_n\left(\frac{x}{l_H}\right), \tag{S.34}$$

where $H_n(x)$ – Hermitian polynomials. Applying time-reversal symmetry, we obtain the eigenfunctions for holes:

$$\psi_{n,-k}^*(x,y) = \frac{1}{\sqrt{L}} e^{iky} \phi_n(x + x_k). \tag{S.35}$$

Without superconducting pairing $\Delta_0 = 0$, the energy spectrum consists of doubly degenerate Landau levels: $E = \pm\omega_c(n+1/2)$. To diagonalize Eq. (S.34), we compute the matrix elements of $\Delta_0$, which can be determined analytically:

$$\langle\psi_{m,-k}^*|\Delta_0|\psi_{n,k'}\rangle = \Delta_0 \delta_{k,k'} \sqrt{\frac{2^{m-n} n!}{m!}} e^{-\xi^2} \xi^{m-n} L_n^{m-n}(2\xi^2), \quad m > n, \tag{S.36}$$

$$\langle \psi^*_{m,-k}|\Delta_0|\psi_{n,k'}\rangle = \Delta_0 \delta_{k,k'} \sqrt{\frac{2^{n-m}m!}{n!}} e^{-\xi^2}(-\xi)^{n-m} L_m^{n-m}(2\xi^2), \quad m<n, \tag{S.37}$$

where $L_n^k(x)$ are generalized Laguerre polynomials, and $\xi = x_k/l_H = kl_H$. By numerically diagonalizing Eq. (S.34) for four Landau levels, we obtain the quasiparticle spectrum as a function of the wave vector $k$. A key feature of this spectrum is the dispersion of Landau levels at small wave vectors $kl_H \ll 1$, where the matrix elements of $\Delta_0$ are maximal (see left panel Fig. 1).

Next, we analyze a system confined within a strip of width $W$, following the approach of B. Halperin [2]. We impose boundary conditions such that both electron and hole wavefunctions vanish at the edges of the strip: $\psi_{n,k}(-W/2,y) = \psi_{n,k}(W/2,y) = 0$. A convenient representation for the electron wavefunctions is:

$$\psi_{s,k}(x,y) = \frac{N_{s,k}e^{iky}}{\sqrt{L}}\left[\frac{\mathcal{D}_s\left(\frac{x-x_k}{l_H/\sqrt{2}}\right)}{\mathcal{D}_s\left(-\frac{W/2+x_k}{l_H/\sqrt{2}}\right)} - \frac{\mathcal{D}_s\left(-\frac{x-x_k}{l_H/\sqrt{2}}\right)}{\mathcal{D}_s\left(\frac{W/2+x_k}{l_H/\sqrt{2}}\right)}\right], \tag{S.38}$$

where $\mathcal{D}_s(x)$ is the parabolic cylinder function with a continuous index $s$, , determined self-consistently, and $N_{s,k}$ is a normalization constant. The boundary condition at $\psi_{n,k}(-W/2,y) = 0$ is automatically satisfied, while the condition at $x = W/2$ leads to the transcendental equation for $s$:

$$\frac{\mathcal{D}_s\left(\frac{W/2-x_k}{l_H/\sqrt{2}}\right)}{\mathcal{D}_s\left(-\frac{W/2+x_k}{l_H/\sqrt{2}}\right)} = \frac{\mathcal{D}_s\left(-\frac{W/2-x_k}{l_H/\sqrt{2}}\right)}{\mathcal{D}_s\left(\frac{W/2+x_k}{l_H/\sqrt{2}}\right)}. \tag{S.39}$$

Solving Eq. (S.39) determines $s(k)$. The spectrum in the absence of superconducting pairing follows a Landau-like form: $E = \pm\omega_c(s(k) + 1/2)$. For a nonzero pairing potential, we employ perturbation theory under the assumption $\Delta_0 \ll \omega_c$. This allows us to restrict the Hamiltonian to the first Landau level (the lowest root of Eq. (S.39)). The effective Hamiltonian is:

$$\mathcal{H} = \begin{pmatrix} \omega_c(s(k)+1/2) & \langle\psi^*_{0,-k}|\Delta_0|\psi_{0,k'}\rangle \\ \langle\psi_{0,k}|\Delta_0|\psi^*_{0,-k'}\rangle & -\omega_c(s(-k)+1/2) \end{pmatrix}. \tag{S.40}$$

To diagonalize this Hamiltonian, we first solve Eq.(S.39) numerically and then compute the matrix element $\langle\psi^*_{0,-k}|\Delta_0|\psi_{0,k'}\rangle$ numerically as well (see Fig.1). The resulting quasiparticle spectrum is shown in Fig.1 from main text.

### S.III. STRUCTURE OF THE FLUCTUATION PROPAGATORS WITH ZEEMAN FIELD.

In this section, we compute diffusive propagators in the presence of a non-zero Zeeman field $B_z$, considering only the RA block of Matsubara space. For this purpose, we consider the NL$\sigma$M action for class C with a symmetry-breaking term:

$$S_{\rm b} = -\frac{g}{16}\int_{\boldsymbol{x}} \mathrm{Tr}\,(\nabla Q)^2 + i\pi\nu\mu_B B_Z \int d^2\boldsymbol{x}\, \mathrm{Tr}\,\mathsf{s}_3 Q. \tag{S.41}$$

Here, we omit the topological term because there are no perturbative corrections to it. It is convenient to parametrize the $Q$-matrix in terms of fluctuation matrices $W$, using the root parametrization:

$$Q = W + \Lambda\sqrt{1-W^2}, \quad W = \begin{pmatrix} 0 & w \\ w^\dagger & 0 \end{pmatrix}, \tag{S.42}$$

From the BdG symmetry relation, one can obtain restrictions on the Nambu components of matrices $w = w_i \mathsf{s}_i$:

$$w_i^{\alpha\beta} = v_i w_i^{\beta\alpha}, \quad v_i = \{-1,1,1,1\}. \tag{S.43}$$

Therefore, the singlet ($i=0$) part of $w$ is a skew-symmetric matrix, while the triplet ($i \neq 0$) parts are symmetric. Expanding the action (S.41) to second order in $W$, we obtain the quadratic action for fluctuations:

$$S^{(2)} = \int_{\boldsymbol{x}} w_i^{\alpha\beta}\left(\frac{g}{4}\delta_{ij}\nabla^2 - i\pi\mu_B\nu B_z M_{ij}\right) w_j^{\dagger\beta\alpha}. \tag{S.44}$$

Here, $M_{ij} = \text{tr}_N(\mathsf{s}_3[\mathsf{s}_i, \mathsf{s}_j]) = 4i(\delta_{i1}\delta_{j2} - \delta_{i2}\delta_{j1})$ and summation over coinciding indices is implied. Transforming to Fourier space:

$$S^{(2)} = -\int_{\boldsymbol{q}} w_i^{\alpha\beta}(\boldsymbol{q}) \left(\frac{g}{4}\delta_{ij}q^2 + i\pi\mu_B \nu B_z M_{ij}\right) w_j^{\dagger\beta\alpha}(-\boldsymbol{q}). \tag{S.45}$$

To invert this matrix, we use the symmetry relations for $w$ matrices (S.43). We split the action into two parts: diagonal and non-diagonal in replica space:

$$w_i^{\alpha\beta}\delta_{ij}w_j^{\dagger\beta\alpha} = \frac{1}{2}\sum_{\alpha=1}^{N_r} w_i^{\alpha\alpha}\delta_{ij}(1+v_j)w_j^{\dagger\alpha\alpha} + 2\sum_{1\le\alpha<\beta}^{N_r} w_i^{\alpha\beta}\delta_{ij}v_j w_j^{\dagger\alpha\beta}, \tag{S.46}$$

$$w_i^{\alpha\beta}M_{ij}w_j^{\dagger\beta\alpha} = \sum_{\alpha=1}^{N_r} w_i^{\alpha\alpha}M_{ij}w_j^{\dagger\alpha\alpha} + 2\sum_{1\le\alpha<\beta}^{N_r} w_i^{\alpha\beta}M_{ij}w_j^{\dagger\alpha\beta}. \tag{S.47}$$

This form of matrices leads to diffusive propagators:

$$\left\langle w_i^{\alpha\alpha}(\boldsymbol{q})w_j^{\dagger\alpha\alpha}(-\boldsymbol{q})\right\rangle = \begin{pmatrix} 0 & 0 & 0 & 0 \\ 0 & \mathcal{D}_{B_z} & \mathcal{P}_{B_z} & 0 \\ 0 & -\mathcal{P}_{B_z} & \mathcal{D}_{B_z} & 0 \\ 0 & 0 & 0 & \mathcal{D}_0 \end{pmatrix}_{ij}, \quad \left\langle w_i^{\alpha\beta}(\boldsymbol{q})w_j^{\dagger\mu\nu}(-\boldsymbol{q})\right\rangle = \frac{1}{2}\begin{pmatrix} -\mathcal{D}_0 & 0 & 0 & 0 \\ 0 & \mathcal{D}_{B_z} & \mathcal{P}_{B_z} & 0 \\ 0 & -\mathcal{P}_{B_z} & \mathcal{D}_{B_z} & 0 \\ 0 & 0 & 0 & \mathcal{D}_0 \end{pmatrix}_{ij} \delta^{\alpha\mu}\delta^{\beta\nu}, \tag{S.48}$$

where we define:

$$\mathcal{D}_{B_z} = \frac{4gq^2}{g^2 q^4 + 256(\pi\mu_B\nu B_z)^2}, \quad \mathcal{P}_{B_z} = \frac{64\pi\mu_B\nu B_z}{g^2 q^4 + 256(\pi\nu\mu_B B_z)^2}. \tag{S.49}$$

This form of Gaussian propagators indicates that presence of Zeeman field leads to the generation of mass (finite correlation length) in order of $L_B \sim \sqrt{g/\nu\mu_B B_z}$ for triplet modes $w_1$ and $w_2$. Consequently, in the limit of an infinite Zeeman field $B_z \to \infty$, we should extract these fluctuation fields from the NL$\sigma$M manifold of massless modes. The dimension of the corresponding manifold after such an extraction is:

$$\underbrace{n(n-1)}_{\text{singlet non-instanton}} + \underbrace{3n(n+1)}_{\text{triplet}} - \underbrace{2n(n+1)}_{\text{massive triplet}} = \underbrace{4n^2 + 2n}_{\text{class C non-instanton}} - \underbrace{2n(n+1)}_{\text{massive triplet}} = 2n^2, \tag{S.50}$$

where $n = N/2$. Consequently, number of real parameters equal to $2n^2$ corresponds to class A: $Q \in \text{U}(2n)/\text{U}(n)\times\text{U}(n)$. We conclude that switching-on of infinite Zeeman field split class C ($Q \in \text{Sp}(4n)/\text{U}(2n)$) action into class A action.

## S.IV. SUPPRESSION OF $W = 1$ INSTANTON IN LDOS

In this section, we demonstrate the suppression of $W = 1$ instantons with scale sizes $\lambda > L_B$. From a straightforward instanton calculation (for details, see Ref. [3]), one can obtain the renormalization of the average LDoS (for brevity we put number of replicas $N_r = 1$):

$$\delta\nu_{\text{inst}} \propto \int_\ell^L \frac{d\lambda}{\lambda}\nu g e^{-\pi g}\cos(\pi g_H)\int_0^\infty dr_0 \frac{r_0}{r_0^2 + \lambda^2}. \tag{S.51}$$

Here, $\nu(\lambda)$ and $g(\lambda)$ are renormalized through perturbation theory in the Pauli-Villars regularization scheme. The integral over $r_0$ exhibits a logarithmic divergence at large scales. To handle this, we use a cutoff scheme since we calculate quantity up to numerical constant. In the presence of a Zeeman field, we assume that the instanton-induced measure does not influence physical observables on scales larger than $L_B$. This allows us to explicitly calculate the $r_0$ integral:

$$\delta\nu_{\text{inst}} \propto \int_\ell^L \frac{d\lambda}{\lambda}\nu g e^{-\pi g}\cos(\pi g_H)\log\left(1 + \frac{L_B^2}{\lambda^2}\right). \tag{S.52}$$

From this integral, we derive a condition on $L_B$ that defines the presence of the crossover regime: $L_B < \xi_l^{(C)} \propto \ell e^{\pi g}$. We note, that in the limiting case $L_B \to \ell e^{\pi g}$, the term $\log\left(1 + \frac{L_B^2}{\lambda^2}\right) \sim g$, and we recover the result from the original paper. The appearance of the crossover induced instanton measure leads to three distinct cases based on characteristic scales:

$$\log\left(1 + \frac{L_B^2}{\lambda^2}\right) \approx \begin{cases} 2|\log \frac{L_B}{\lambda}|, & \lambda \ll L_B, \\ \text{const}, & \lambda \sim L_B, \\ \frac{L_B^2}{\lambda^2}, & \lambda \gg L_B. \end{cases} \tag{S.53}$$

We observe that for $\lambda \gg L_B$, the logarithmic divergence in the integral over $\lambda$ disappears. This indicates that instantons with scales $\lambda > L_B$ do not contribute to the renormalization of LDoS.

## S.V. VARIATIONAL PRINCIPLE FOR $W = 2$ SOLUTION

The expression for the Zeeman term in the classical action (equation (A4) in End Matter) suggests that the zero modes of the two-instanton solution transform across the crossover [4]. For brevity, we put $z_0^{(1)} = z_0^{(2)} = 0$. We identify two possible mechanisms for this transformation: (1) synchronization of the instanton scale sizes, $\lambda_1 = \lambda_2$, or (2) the disappearance of the mixing angle $\chi$. At the classical level, we cannot determine which possibility is more favorable based on the least action principle alone.

To resolve this, we adopt a variational ansatz for the zero modes in the following form:

$$\lambda_i(r) = \lambda_i\left(1 + \frac{\lambda_i^* - \lambda_i}{\lambda_i}\tanh\frac{r^2}{\rho^2}\right), \quad \chi(r) = \chi_0 + \chi_1\tanh\frac{r^2}{\rho^2}, \tag{S.54}$$

where the free parameters are $\lambda_i^*$ (the final instanton scale), $\rho$ (the characteristic scale of zero-mode evolution), and $\chi_1$ (the final value of the mixing angle). By substituting this ansatz into the action $S_\text{b}$ (Equation (7) in the main text) and performing numerical minimization, we find that the synchronization of instanton scale sizes is the correct mechanism for instanton zero-mode transformation. The final values for the free parameters are $\lambda_1^* = \lambda_2^* \approx (\lambda_1 + \lambda_2)/2$ and $\chi_1$ does not vanish $\chi_1 \approx \text{const} \neq -\chi_0$.

## S.VI. PLATEAU WIDTH IN THE LIMIT OF STRONG ZEEMAN FIELD.

In this section, we compute the critical shift $\delta g_H$ of the transverse spin conductivity microscopic value, necessary to reach the $g_H = 0$ phase in the infrared limit ($L \to \infty$). To achieve this, we consider the RG equations for class C, including non-perturbative corrections, as obtained in Ref. [3] and which were discussed in End Matter. These equations are given as:

$$\begin{aligned} \frac{dg}{d\ln L} &= -\frac{1}{\pi} - \frac{2}{\pi^2 g} - \mathcal{D}_\text{C}(\pi g)^3 e^{-\pi g}\cos\pi g_H, \\ \frac{dg_H}{d\ln L} &= -\mathcal{D}_\text{C}(\pi g)^3 e^{-\pi g}\sin\pi g_H, \end{aligned} \tag{S.55}$$

where we focus only on the weak-coupling limit $\bar{g} \gg 1$. In this limit, we retain only the one-loop correction in the $g$-equation. Consequently, the system reduces to a trajectory-type dimensionless equation with initial conditions $g(\ell) = \bar{g}$ and $g_H(\ell) = \bar{g}_\text{H}$:

$$\frac{dx}{dy} = \pi\mathcal{D}_\text{C} y^3 e^{-y}\sin x, \tag{S.56}$$

where $x = \pi g_H$ and $y = \pi g$. This equation can be solved to yield:

$$g_H(L) = \frac{2}{\pi}\arctan\left[\tan\left(\frac{\pi\bar{g}_\text{H}}{2}\right)e^{\pi\mathcal{D}_\text{C} f(g(L),\bar{g})}\right], \tag{S.57}$$

where we define:

$$f(g(L),\bar{g}) = \int_{\pi\bar{g}}^{\pi g} y^3 e^{-y} dy, \quad g(L) = \bar{g} - \frac{1}{\pi}\log\frac{L}{\ell}. \tag{S.58}$$

The condition defining the critical trajectory can be expressed as:

$$\frac{1}{2} = \frac{2}{\pi} \arctan \left[ \tan \left( \frac{\pi \bar{g}_\mathrm{H}}{2} \right) e^{\pi \mathcal{D}_\mathrm{C} f(g(L_B), \bar{g})} \right]. \tag{S.59}$$

From this equation, the corresponding starting point $\bar{g}_\mathrm{H}$ can be found:

$$\bar{g}_\mathrm{H} = \frac{2}{\pi} \tan^{-1} \left( e^{-\pi \mathcal{D}_\mathrm{C} f(g(L_B), \bar{g})} \right). \tag{S.60}$$

### S.VII. SOLUTION OF CROSSOVER RG EQUATIONS IN WEAK COUPLING LIMIT $g \gg 1$.

In this section, we solve crossover RG equation. As the crossover occurs sharply, we can approximate the crossover function as a Heaviside step-function (see End Matter):

$$\frac{dg}{d \ln L} = -\frac{1}{\pi} \theta(L_B - L) - \frac{1}{2\pi^2 g} \theta(L - L_B), \tag{S.61}$$

$$\frac{dg_H}{d \ln L} = -\theta(L_B - L) \mathcal{D}_\mathrm{C} (\pi g)^3 e^{-\pi g} \sin \pi g_H - \theta(L - L_B) \mathcal{D}_\mathrm{A} (\pi g)^2 e^{-2\pi g} \sin 2\pi g_H, \tag{S.62}$$

where we neglect non-perturbative corrections in the first equation. Using this, we can solve these equations in a manner similar to the previous section. The solution can be expressed as:

$$g_H(L) = \frac{1}{\pi} \begin{cases} 2 \arctan \left( \tan \left( \pi \bar{g}_\mathrm{H}/2 \right) \exp \left[ \pi \mathcal{D}_C f(g_C(L), \bar{g}) \right] \right), & L \leq L_B \\ \arctan \left( \tan \left( \pi \bar{\bar{g}}_\mathrm{H} \right) \exp \left[ \pi \mathcal{D}_A f(2 g_A(L), 2\bar{\bar{g}}) \right] \right), & L > L_B \end{cases} \tag{S.63}$$

where the function $f$ is defined in Eq. (S.58), and the terms $g_C(L), g_A(L)$ are longitudinal spin conductivities with one(two)-loop corrections in symmetry class C (A):

$$g_C(L) = \bar{g} - \log \frac{L}{\ell}, \quad g_A(L) = \left( \bar{\bar{g}}^2 - \frac{1}{\pi} \log \frac{L}{L_B} \right)^{1/2}. \tag{S.64}$$

To determine the period of $g_H(L)$ as a function of $\bar{g}_\mathrm{H}$, we simplify the trigonometric terms in the solution, yielding:

$$g_H(L) = \frac{1}{\pi} \arctan \left( \frac{2 \sin(\pi \bar{g}_\mathrm{H})}{1 + \cos(\pi \bar{g}_\mathrm{H}) \frac{\exp 2\mathcal{F}_\mathcal{C} + 1}{1 - \exp 2\mathcal{F}_\mathcal{C}}} \frac{\exp \left( \mathcal{F}_\mathcal{C} + \mathcal{F}_\mathcal{A} \right)}{1 - \exp 2\mathcal{F}_\mathcal{C}} \right), \tag{S.65}$$

where we define $\mathcal{F}_i = \pi \mathcal{D}_i f(g(L), g(L_0))$. From this expression, it follows that when $L_B > C \xi_l^{(C)}$, with numerical constant $C \approx 1.67 \times 10^{-3}$, the trigonometric part simplifies to $\tan \pi \bar{g}_\mathrm{H}$, leading to quantization of transverse spin conductivity by integers. Conversely, when $L_B < C \xi_l^{(C)}$, the quantization changes to even integers. This result is in full agreement with the logic expressed in the main text.

---



\end{document}